\documentclass[twocolumn,prl,showpacs,superscriptaddress,preprintnumbers,amsmath,amssymb]{revtex4}
\newcommand{\CoMn}{Ca$_3$CoMnO$_6$}

\usepackage[dvips]{graphicx}
\usepackage{dcolumn}
\usepackage{bm}
\begin{document}

\title{Ising magnetism and ferroelectricity in Ca$_3$CoMnO$_6$}

\author{Hua Wu}
\affiliation{II. Physikalisches Institut, Universit{\"a}t zu K{\"o}ln,
  Z{\"u}lpicher Str. 77, 50937 K{\"o}ln, Germany}
\author{T. Burnus}
\affiliation{II. Physikalisches Institut, Universit{\"a}t zu K{\"o}ln,
  Z{\"u}lpicher Str. 77, 50937 K{\"o}ln, Germany}
\author{Z. Hu}
\affiliation{II. Physikalisches Institut, Universit{\"a}t zu K{\"o}ln,
  Z{\"u}lpicher Str. 77, 50937 K{\"o}ln, Germany}
\author{C. Martin}
\affiliation{Laboratoire CRISMAT, UMR 6508 CNRS ENSICAEN, 14050 Caen, France}
\author{A. Maignan}
\affiliation{Laboratoire CRISMAT, UMR 6508 CNRS ENSICAEN, 14050 Caen, France}
\author{J. C. Cezar}
\affiliation{European Synchrotron Radiation Facility, Bo\^ite Postale 220,
38043 Grenoble C\'edex, France}
\author{A. Tanaka}
\affiliation{Department of Quantum Matter, ADSM, Hiroshima University,
Higashi-Hiroshima 739-8530, Japan}
\author{N. B. Brookes}
\affiliation{European Synchrotron Radiation Facility, Bo\^ite Postale 220,
38043 Grenoble C\'edex, France}
\author{D. I. Khomskii}
\affiliation{II. Physikalisches Institut, Universit{\"a}t zu K{\"o}ln,
  Z{\"u}lpicher Str. 77, 50937 K{\"o}ln, Germany}
\author{L. H. Tjeng}
\affiliation{II. Physikalisches Institut, Universit{\"a}t zu K{\"o}ln,
  Z{\"u}lpicher Str. 77, 50937 K{\"o}ln, Germany}

\date{\today}

\begin{abstract}
The origin of both the Ising chain magnetism and ferroelectricity
in Ca$_3$CoMnO$_6$ is studied by $ab$ $initio$ electronic
structure calculations and x-ray absorption spectroscopy. We find
that Ca$_3$CoMnO$_6$ has the alternate trigonal prismatic
Co$^{2+}$ and octahedral Mn$^{4+}$ sites in the spin chain. Both
the Co$^{2+}$ and Mn$^{4+}$ are in the high spin state. In
addition, the Co$^{2+}$ has a huge orbital moment of 1.7 $\mu_B$
which is responsible for the significant Ising magnetism.
The centrosymmetric crystal structure known so far is
calculated to be unstable with respect to exchange striction in the
experimentally observed $\uparrow\uparrow\downarrow\downarrow$
antiferromagnetic structure for the Ising chain. The calculated
inequivalence of the Co-Mn distances accounts for the
ferroelectricity.
\end{abstract}

\pacs{71.20.-b, 78.70.Dm, 71.70.-d, 71.27.+a}

\maketitle

Among a variety of multiferroic materials discovered
so far \cite{Cheong07,Khomskii},
ferroelectric Ising chain magnet {\CoMn} is quite unique, because the
ferroelectricity (FE)
is driven by exchange striction in a collinear Ising spin chain consisting
of the charge ordered transition-metal ions \cite{Choi08}. The spin chain
has the alternate trigonal prismatic and octahedral
sites \cite{Choi08,Zubkov01}.
Special $\uparrow\uparrow\downarrow\downarrow$
antiferromagnetic (AF) structure is detected in {\CoMn}
below T$_{\rm N}\approx$13 K by neutron diffraction. However, the measured magnetic
moment of 0.66 $\mu_B$/Co and 1.93 $\mu_B$/Mn is much smaller
than the expected one of the normal high-spin (HS) Co$^{2+}$ ($S$=3/2) and
Mn$^{4+}$ ($S$=3/2). This led Choi $et$ $al.$ to a conclusion that the
Co$^{2+}$ is (surprisingly) in a low-spin (LS) state \cite{Choi08}.
In contrast, the effective magnetic moment of $\mu_{\rm eff}$=5.8-6.0 $\mu_B$
per formula unit (f.u.),
extracted from magnetic susceptibility measurements
above T$_{\rm N}$ \cite{Zubkov01,Rayaprol03},
suggests that both Co$^{2+}$ and Mn$^{4+}$ are in a HS state.
Thus there is an apparent controversy
between those data, and the problem concerning the spin state of,
in particular, Co$^{2+}$ ions seems to be still unsolved.
Another important question is to understand the nature of the
Ising magnetism and of the resulting
exchange striction, which are apparently crucial for the appearance of FE in
Ca$_3$CoMnO$_6$.
To this end, we carried out $ab$ $initio$ electronic structure calculations
and x-ray absorption spectroscopy (XAS). We address
the important issues including the Co/Mn site preference, their
charge/spin/orbital states, the origin of the Ising magnetism, and
the exchange striction.

Our $ab$ $initio$ calculations were performed by using the full-potential
augmented plane waves plus local orbital method (Wien2k code) \cite{WIEN2k}. We
took the experimental centrosymmetric structure data of the rhombohedral lattice
($R$-3$c$) which has in a hexagonal setting the lattice constant
$a$=9.1314 \AA~and $c$=10.5817 \AA~\cite{Zubkov01,Kiryukhin}.
The calculations were done for different magnetic structures
($\uparrow\downarrow\uparrow\downarrow$,
$\uparrow\uparrow\downarrow\downarrow$,
$\uparrow\downarrow\downarrow\downarrow$,
and $\uparrow\uparrow\uparrow\uparrow$ orderings in the Co-Mn-Co-Mn chains).
To study the exchange striction effect, we investigated the effect of
internal atomic relaxation
allowing the inversion symmetry to be broken in the
$\uparrow\uparrow\downarrow\downarrow$ spin chain,
as discussed below.
The muffin-tin sphere radii are chosen to be 2.4, 2.1 and 1.4 Bohr
for Ca, Co/Mn and O atoms, respectively. The cut-off energy of 16
Ryd is used for plane wave expansion,
and 5$\times$5$\times$5 {\bf k}-mesh for integrations
over the Brillouin zone.
To account for the strong electron correlations,
GGA+U (the generalized gradient approximation \cite{Perdew}
plus Hubbard $U$) calculations \cite{Anisimov93} were performed.
$U$=5.0 (4.0) eV for the Co (Mn) $3d$ electrons (with a common Hund exchange
of 0.9 eV) are used, according to the calculations using a local-orbital
basis \cite{Pickett98}. Note also that the choices of $U$ value in the range of
2-7 eV do not affect the conclusion made in this Letter.
The spin-orbit coupling (SOC) turns out to be crucial
and it is included by the second-variational method with scalar relativistic
wave functions \cite{WIEN2k}.

\begin{figure}[ht]
 \centering\includegraphics[width=6.5cm]{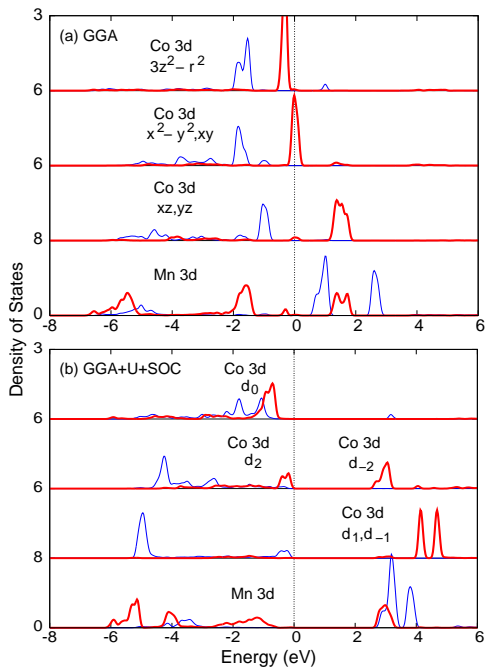}
\vskip-0.2cm
 \caption{(Color online) Density of states calculated by GGA (a) and GGA+U+SOC (b)
for the $\uparrow\downarrow\uparrow\downarrow$ spin structure of {\CoMn} with the
trigonal Co and octahedral Mn. It practically coincides with that for
$\uparrow\uparrow\downarrow\downarrow$.
The thin blue (bold red) curves refer to the majority (minority) spin. The Fermi
level is set at zero energy. In (a) the Mn$_{\rm oct}^{4+}$ has a closed $t_{2g}^3$ shell
centered at --1.5 eV, together with the $e_g$ bonding state around --5.5 eV.
The Co$_{\rm trig}$ is in the high-spin 2+ state with
the nearly degenerate $3z^2$--$r^2$ ($d_0$), $x^2$--$y^2$
and $xy$ ($d_{\pm 2}$) levels.
In (b) the ($x^2$--$y^2$, $xy$) doublet splits due to
the spin-orbit coupling,
and the Hubbard $U$ yields an insulating ground state with the minority-spin
$d_0d_2$ occupation.}
\vskip-0.35cm
 \label{fig1}
\end{figure}

It is quite common that an octahedral Co$^{3+}$ ion is in a LS ground state.
This may suggest that the small moment of 0.66 $\mu_B$/Co in {\CoMn}
could be due to the Co/Mn site disorder, i.e., an appreciable
presence of the octahedral LS Co$^{3+}$. To probe the Co/Mn site preference,
we computed two structures either with the trigonal Co (Co$_{\rm trig}$) and
octahedral Mn (Mn$_{\rm oct}$)
or vice versa. The total energy results show that the former structure
is more stable than the later by 0.33 eV/f.u. in GGA and more significantly,
by 1.60 eV/f.u. in GGA+U+SOC. The energetically favorable structure has
the HS trigonal Co$^{2+}$ and the HS octahedral Mn$^{4+}$, while the unfavorable
one indeed has the LS octahedral Co$^{3+}$ and the HS trigonal Mn$^{3+}$.
We plot in Fig. 1 the
GGA and GGA+U+SOC calculated density of states (DOS) only for the favorable
structure. The sharp DOS peak at the Fermi level, coming from the degenerate
$x^2$--$y^2$ and $xy$ ($d_{\pm 2}$) levels of the trigonal Co$^{2+}$
minority-spin
$d$ electrons, vanishes when going from the GGA to GGA+U+SOC solutions.
This explains why, by opening a sizable gap of 2.4 eV, the GGA+U+SOC strongly favors
the structure with the Co$^{2+}_{\rm trig}$ and Mn$^{4+}_{\rm oct}$.
The HS Co$^{2+}$ (Mn$^{4+}$) has a calculated spin moment of 2.64 (2.70) $\mu_B$,
both staying constant within 0.2 $\mu_B$ for $U$=2-7 eV.

Having established the Co$_{\rm trig}^{2+}$/Mn$_{\rm oct}^{4+}$ site preference from the above
total-energy calculations, we turn to our XAS measurements to confirm it.
The room temperature Co-$L_{2,3}$ and Mn-$L_{2,3}$ XAS of
{\CoMn} were collected at the ID08 beamline of the
European Synchrotron Radiation Facility (ESRF) in Grenoble with a
resolution of 0.25 eV at Co-$L_3$ (at 780 eV). The sharp peak at
777.8~eV of the Co-$L_3$ edge of single crystalline CoO and at
640~eV of the Mn-$L_3$ of single crystalline MnO were used for
energy calibration. The spectra were recorded using the total
electron yield method by measuring the sample drain current in a
chamber with a base pressure of $2\times10^{-10}$ mbar. Clean
sample areas were obtained by cleaving the polycrystals \textit{in
situ}.

\begin{figure}[ht!]
 \centering\includegraphics[width=6cm]{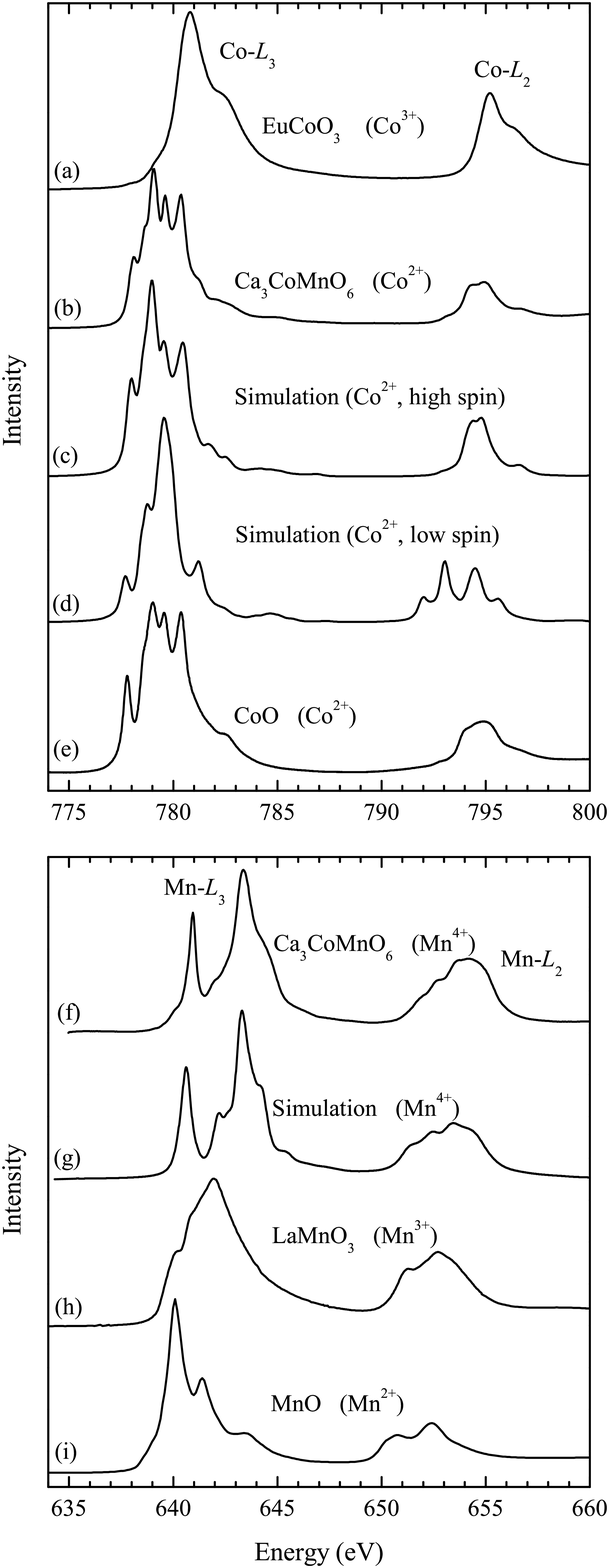}
\vskip-0.2cm
 \caption{The experimental and simulated x-ray absorption spectra of {\CoMn} at
the Co-$L_{2,3}$ (upper panel) and the Mn-$L_{2,3}$ (lower panel) edges, together
with the experimental spectra of the references CoO, EuCoO$_3$, MnO, and LaMnO$_3$.
{\CoMn} turns out to have the high-spin trigonal Co$^{2+}$ and octahedral
Mn$^{4+}$.}
\vskip-0.4cm
 \label{fig2}
\end{figure}

Important is that XAS spectra are highly sensitive to the valence
state: an increase of the valence state of the transition metal ion by one
causes a shift of the XAS $L_{2,3}$ spectra by one or more eV
towards higher energies \cite{Mitra03,Burnus08}. In
Fig. 2 (upper panel) we see a shift of the center of gravity
of the Co-$L_3$ white line to higher photon energies by about
1.5 eV in going from the divalent CoO to the trivalent EuCoO$_3$.
The energy position of the Co line in Ca$_3$CoMnO$_6$ is the same as in CoO,
indicating a Co$^{2+}$ state. Also, a gradual shift of the center of
gravity of the Mn-$L_3$ white line to higher energies from MnO
via LaMnO$_3$ to Ca$_3$CoMnO$_6$ (lower panel in Fig. 2) evidences the
increase of the Mn valence state from $2+$ via $3+$ to
$4+$. In fact the Mn-$L_{2,3}$ edges of Ca$_3$CoMnO$_6$ lie at
the same energy position as in SrMnO$_3$,
LaMn$_{0.5}$Co$_{0.5}$O$_{3}$ \cite{Burnus08}, and
LaMn$_{0.5}$Ni$_{0.5}$O$_{3}$ \cite{Sanchez02}, all of which
have the Mn$^{4+}$ state.
Thus, our XAS results confirm the Co$^{2+}$-Mn$^{4+}$ state.

To extract more detailed information concerning the charge and
spin states from the Co-$L_{2,3}$ and the Mn-$L_{2,3}$ XAS
spectra, we have carried out simulations of the XAS spectra
using the well-proven configuration-interaction cluster
model \cite{Tanaka94,deGroot94,Thole97}. We studied
a trigonal prism CoO$_6$ (an octahedral MnO$_6$) cluster, including
the full atomic multiplet theory and the local effects of the
solid. Thus our model calculations account for the intra-atomic $3d$--$3d$ and $2p$--$3d$
Coulomb interactions, the atomic $2p$ and $3d$ spin-orbit
couplings, the O $2p$--Co $3d$ hybridization, and the proper local
crystal-field parameters. The calculated
Co-$L_{2,3}$ XAS spectrum for the HS Co$^{2+}$ [curve (c) in Fig. 2]
with the ionic trigonal crystal
field interaction  $\Delta_{10}^{\rm ionic}$=0.75 eV
reproduces the experimental spectrum [curve (b)] very
well \cite{parameter}. In order to stabilize a LS state which
was concluded by Choi $et$ $al$ \cite{Choi08}, we would have to
increase the $\Delta_{10}^{\rm ionic}$ by nearly 3 times
(2.8 eV), and the calculated spectrum for this case
[curve (d) in Fig. 2] strongly
disagrees with the experimental one [curve (b)].

In the lower panel of Fig. 2, one can see that the lineshapes of the
experimental Mn spectrum [curve (f)] are well reproduced by the
simulation [curve (g)] with a Mn$^{4+}$ ($t_{2g}^3$) configuration
in a local O$_h$ symmetry.

Thus, our $ab$ $initio$ calculations and XAS experiments
have firmly established that the spin-chain magnet {\CoMn} has the HS
Co$^{2+}_{\rm trig}$ and HS Mn$^{4+}_{\rm oct}$, and that there is no
appreciable presence of the LS-Co$^{3+}_{\rm oct}$/HS-Mn$^{3+}_{\rm trig}$.
Obviously, the magnetic moment of 0.66 $\mu_B$/Co measured by neutron
diffraction \cite{Choi08} below T$_{\rm N}\approx$13 K is much smaller than our
theoretical value [the total calculated moment being about 4.3 $\mu_B$/Co with the
spin (orbital) contribution of 2.6 (1.7) $\mu_B$, see more below]
and than the value of $\mu_{\rm eff}$=5.8-6.0 $\mu_B$ obtained from the
high-temperature magnetic
susceptibility \cite{Zubkov01,Rayaprol03}. This may be partially due to
strong fluctuations in quasi one-dimensional chains, with frustrated interchain
interactions. Another important factor may be a close proximity of different
types of magnetic orderings, see below. This question deserves further study.

\begin{figure}[h]
 \centering\includegraphics[width=6cm]{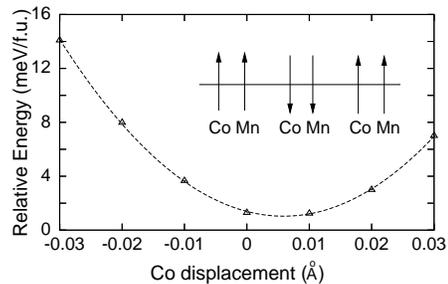}
\vskip-0.2cm
 \caption{Relative energy as function of the Co displacements calculated by GGA+U+SOC for
the experimental lattice. Triangles stand for the data points, which are fitted into
a parabolic curve. The energy minimum shows the Co displacement of 0.006 \AA, which leads
to alternate short-long-short-long Co-Mn distances (the difference being
0.012 \AA)
in the Co$\uparrow$-Mn$\uparrow$-Co$\downarrow$-Mn$\downarrow$ chain as depicted schematically
in inset.}
\vskip-0.35cm
 \label{fig3}
\end{figure}

As seen in Fig. 1(a), the trigonal Co$^{2+}$ has the nearly degenerate
$3z^2$--$r^2$ and ($x^2$--$y^2$, $xy$) levels. Due to the in-plane character
of both the $x^2$--$y^2$ and $xy$ orbitals, their strong Coulomb repulsion
prevents their double occupation in the minority-spin channel. Therefore,
the minority-spin $3z^2$--$r^2$ orbital is fully occupied and the minority-spin
$x^2$--$y^2$ and $xy$ are half-filled for the HS Co$^{2+}$ ions.
Due to the quasi one-dimensional character along the $c$-axis chain,
a naive $x^2$--$y^2$/$xy$ planar orbital ordering does not gain any energy
(compared with either an $x^2$--$y^2$/$x^2$--$y^2$ or $xy$/$xy$ orbital
ordering), as proved by
our $ab$ $initio$ calculations. In contrast, an efficient way to gain
the energy is SOC. When the SOC is included, the ($x^2$--$y^2$, $xy$) doublet
splits into $d_2$ and $d_{-2}$, and the gain of the full SOC energy
is calculated to be about 70 meV. As a result, a huge orbital magnetic
moment of 1.7 $\mu_B$ is generated at the HS Co$^{2+}_{\rm trig}$ sites with the
minority-spin $d_0d_2$ occupation [Fig. 1(b)], and the SOC firmly fixes the total
magnetization (with parallel spin and orbital contributions)
along the $c$-axis chain direction. Note that the orbital moment stays constant
within 0.1 $\mu_B$ in our GGA+U+SOC calculations for $U$=2-7 eV.
Therefore, the peculiar
trigonal crystal field and the SOC are responsible for the significant
Ising character of {\CoMn}; cf. similar situation in the isostructural
Ca$_3$CoRhO$_6$ \cite{Wu07}.

Having established the picture about the Co/Mn site preference,
their charge/spin/orbital states and the significant Ising
magnetism, we turn now to the discussion on the intra-chain
magnetism and the exchange striction leading to ferroelectricity.
Since a centrosymmetric structure of {\CoMn} does not
carry ferroelectricity, we now look whether the symmetry can be
lowered by allowing for atomic displacements within the
\textit{experimentally observed} \cite{Choi08}
$\uparrow\uparrow\downarrow\downarrow$ magnetic structure of the
Co-Mn-Co-Mn Ising spin chain: the Co-Mn bonds would be
inequivalent in such a non-centrosymmetric structure leading to
chain dimerization, and this will give rise to ferroelectricity. We
calculated using GGA+U+SOC the total energy of the system by
firstly shifting the Co ions along the chain making Co-Mn
distances unequal (alternating) \cite{note2}. The results are
shown in Fig. 3. We see that indeed the lattice in
$\uparrow\uparrow\downarrow\downarrow$ structure would relax to a
state with alternating Co-Mn distances (the difference being
0.012 \AA), which means the appearance of FE in the system, cf.
\cite{Efremov,Picozzi,Xiang,Brink}. The calculated spin and
orbital moments remain unchanged. Furthermore, a triclinic
lattice ($P$1) was tested to study a likely breaking of the
3-fold rotation symmetry. A structural relaxation using GGA+U+SOC
(with the Co ions fixed at the above optimized position)
confirms the small FE distortion of the Co-Mn chains and gives
also a small distortion of the Co-O (Mn-O) bonds [only about 
0.01 \AA~out of 2.15 \AA~(1.92 \AA) in bondlength]. We would
like to point out that those small distortions are not
inconsistent with the experiment \cite{Kiryukhin} since they are so
small that they would require a special effort for their detection.

We note that in the GGA+U+SOC calculations the
$\uparrow\uparrow\downarrow\downarrow$ and 
$\uparrow\downarrow\uparrow\downarrow$
magnetic structures for the Co-Mn-Co-Mn spin chain 
are the two lowest-energy states, with the former being higher 
in energy. The energy difference is
15 meV/f.u. for the experimentally detected crystal structure
\cite{Zubkov01} and is reduced to 7 meV/f.u. for the 
relaxed triclinic lattice ($P$1). At the moment it is not clear
how to stabilize the experimentally observed
$\uparrow\uparrow\downarrow\downarrow$ magnetic structure
\cite{Choi08} in the GGA+U+SOC calculations. Since the energy
difference is rather small and the intrachain exchange
interactions rather weak, we speculate that the couplings between
the chains may play an important role. The present calculations
assumed a simple ferromagnetic interchain interaction, but it is known that
there are AF interchain couplings in the isostructural
Ca$_3$Co$_2$O$_6$ \cite{Kageyama97} and Ca$_3$CoRhO$_6$
\cite{Niitaka01} compounds with strengths not much smaller than
the intrachain ones. These finite AF interchain couplings bring
about magnetic frustration, and one needs to investigate a large set
of intra- and interchain magnetic structures in order to explain
theoretically the magnetic ground state of {\CoMn}.

In summary, we find the site preference of the high-spin trigonal
Co$^{2+}$ and the octahedral Mn$^{4+}$ in the Ising chain magnet
{\CoMn}, using the $ab$ $initio$ calculations and x-ray absorption
spectroscopy. The Co$^{2+}$ has the very stable minority-spin
$d_0d_2$ occupation due to the peculiar trigonal crystal field,
inter-orbital Coulomb repulsion and the spin-orbit coupling.
Thus, a huge orbital moment of 1.7 $\mu_B$ is predicted and the
significant Ising magnetism is well accounted for. Moreover, our
calculations indeed found that a structural relaxation due to the
exchange striction decreases the energy of the 
experimentally observed $\uparrow\uparrow\downarrow\downarrow$
magnetic ordering and leads to the observed ferroelectricity.
However, the mechanism of the full stabilization
of the $\uparrow\uparrow\downarrow\downarrow$ structure as the
ground state calls for a further study.

We are grateful to V. Kiryukhin and S. W. Cheong for informing us of their
unpublished structural data.
This research was supported by the Deutsche Forschungsgemeinschaft through SFB 608.


\begin{thebibliography}{15}
\bibitem{note2} Co ions are in the relatively open trigonal prisms, compared with
Mn in the octahedra. Thus, Co is more susceptible to the exchange striction, as
proved by the structural relaxation.
\end{thebibliography}
\end{document}